\documentclass[aps,prab,amsmath,amssymb,superscriptaddress,twocolumn,]{revtex4-2}
\usepackage[utf8]{inputenc}
\usepackage{graphicx}
\usepackage{hyperref}
\hypersetup{colorlinks=False,}
\usepackage{multirow}

\begin{document}

\title{High-Power Polarization-Controlled Attosecond-Scale Soft X-ray Pulses}
\author{Zhaoheng Guo}
\email{zhaoheng.guo@psi.ch}
\affiliation{PSI Center for Photon Science, Paul Scherrer Institute, 5232 Villigen PSI, Switzerland}
\affiliation{Laboratory for Ultrafast X-ray Sciences, Institute of Chemical Sciences and Engineering, \'Ecole Polytechnique F\'ed\'erale de Lausanne, 1015 Lausanne, Switzerland}
\author{Carlo Vicario}
\affiliation{PSI Center for Photon Science, Paul Scherrer Institute, 5232 Villigen PSI, Switzerland}
\author{Andre Al Haddad}
\affiliation{PSI Center for Photon Science, Paul Scherrer Institute, 5232 Villigen PSI, Switzerland}
\author{Christopher Arrell}
\affiliation{PSI Center for Photon Science, Paul Scherrer Institute, 5232 Villigen PSI, Switzerland}
\author{Andreas Dax}
\affiliation{PSI Center for Photon Science, Paul Scherrer Institute, 5232 Villigen PSI, Switzerland}
\author{Umit Demirbas}
\affiliation{PSI Center for Photon Science, Paul Scherrer Institute, 5232 Villigen PSI, Switzerland}
\author{Nicole Hiller}
\affiliation{PSI Center for Accelerator Science and Engineering, Paul Scherrer Institute, 5232 Villigen PSI, Switzerland}
\author{Martin Huppert}
\affiliation{PSI Center for Photon Science, Paul Scherrer Institute, 5232 Villigen PSI, Switzerland}
\author{Christoph Kittel}
\affiliation{PSI Center for Accelerator Science and Engineering, Paul Scherrer Institute, 5232 Villigen PSI, Switzerland}
\author{Gregor Knopp}
\affiliation{PSI Center for Photon Science, Paul Scherrer Institute, 5232 Villigen PSI, Switzerland}
\author{Eduard Prat}
\affiliation{PSI Center for Accelerator Science and Engineering, Paul Scherrer Institute, 5232 Villigen PSI, Switzerland}
\author{Sven Reiche}
\affiliation{PSI Center for Accelerator Science and Engineering, Paul Scherrer Institute, 5232 Villigen PSI, Switzerland}
\author{Antoine Sarracini}
\affiliation{PSI Center for Photon Science, Paul Scherrer Institute, 5232 Villigen PSI, Switzerland}
\author{Thomas Schietinger}
\affiliation{PSI Center for Accelerator Science and Engineering, Paul Scherrer Institute, 5232 Villigen PSI, Switzerland}
\author{Kirsten Schnorr}
\affiliation{PSI Center for Photon Science, Paul Scherrer Institute, 5232 Villigen PSI, Switzerland}
\author{Alexandre Trisorio}
\affiliation{PSI Center for Photon Science, Paul Scherrer Institute, 5232 Villigen PSI, Switzerland}
\author{Didier Voulot}
\affiliation{PSI Center for Accelerator Science and Engineering, Paul Scherrer Institute, 5232 Villigen PSI, Switzerland}
\author{Tobias Weilbach}
\affiliation{PSI Center for Accelerator Science and Engineering, Paul Scherrer Institute, 5232 Villigen PSI, Switzerland}
\author{Hankai Zhang}
\affiliation{PSI Center for Photon Science, Paul Scherrer Institute, 5232 Villigen PSI, Switzerland}
\affiliation{Laboratory for Ultrafast X-ray Sciences, Institute of Chemical Sciences and Engineering, \'Ecole Polytechnique F\'ed\'erale de Lausanne, 1015 Lausanne, Switzerland}
\author{Christoph Bostedt}
\affiliation{PSI Center for Photon Science, Paul Scherrer Institute, 5232 Villigen PSI, Switzerland}
\affiliation{Laboratory for Ultrafast X-ray Sciences, Institute of Chemical Sciences and Engineering, \'Ecole Polytechnique F\'ed\'erale de Lausanne, 1015 Lausanne, Switzerland}
\author{Philipp Dijkstal}
\email{philipp.dijkstal@psi.ch}
\affiliation{PSI Center for Accelerator Science and Engineering, Paul Scherrer Institute, 5232 Villigen PSI, Switzerland}

\date{\today}

\begin{abstract}
We demonstrate a versatile platform for high-power attosecond soft X-ray pulse generation with polarization and photon energy control at the SwissFEL free-electron laser. 
An isolated high-current spike embedded within a long electron-beam pedestal emits soft X-ray pulses with single-spike spectra and multi-electronvolt bandwidths in the tunable magnetic fields of Apple-X undulators. 
Demonstrated pulse parameters include a photon energy range of 450--1070~eV, circular as well as linear polarization, and pulse energies from tens to above hundred microjoules. 
By tuning the longitudinal slice-dependent transverse electron beam orbit we can rapidly switch between attosecond and few femtosecond pulse length.
By exploiting magnetic chicanes in the undulator line we can produce two-colour pulse pairs with tunable delay or increase the pulse energy beyond 200~\textmu J through multi-stage amplification schemes. 
High-resolution longitudinal phase-space measurements and start-to-end simulations in addition to spectral measurements provide consistent evidence for attosecond-scale pulse durations. 
This unique combination of high pulse energy and polarization control of attosecond-scale soft X-ray pulses enables the element-specific investigations of spin and chiral dynamics on the natural time scale of electron motion.
\end{abstract}

\maketitle

\section{Introduction}\label{sec:Introduction}

Understanding how electrons move, interact, and reorganize in matter requires experimental tools that match the relevant timescales and symmetries. 
Attosecond soft X-ray pulses enable element-specific direct tracking of charge transfer, chemical bonding, and structural change in molecular, condensed-matter, and nanoscale systems~\cite{krausz2009attosecond, corkum2007attosecond}.
When combined with polarization control, attosecond soft X-rays can also probe spin, orbital, and chiral degrees of freedom, as circularly polarized photons couple directly to these quantities through techniques such as X-ray magnetic circular dichroism~\cite{stohr2006magnetism} and polarization-resolved photoelectron spectroscopy~\cite{beaulieu2017attosecond}. 
Bright, circularly polarized attosecond soft X-rays therefore provide opportunities for studying ultrafast electron dynamics in chiral molecules~\cite{han2025attosecond}, ultrafast magnetism~\cite{siegrist2019light}, and symmetry-dependent phenomena in topological materials~\cite{schmid2021tunable} on their natural timescales with element specificity.

Table-top attosecond light sources driven by high-harmonic generation (HHG) continue to progress, having reached soft X-ray photon energies with linear polarization~\cite{popmintchev2012bright} and ultraviolet photon energies with circular polarization~\cite{kfir2015generation}. 
However, generating circularly polarized attosecond soft X-ray pulses with sufficient photon flux for nonlinear X-ray spectroscopy~\cite{young2010femtosecond,glover2012x,al2022observation} requires a different approach. 
X-ray free-electron lasers (X-ray FELs, or XFELs), driven by relativistic electron beams interacting with the magnetic fields of an undulator beamline~\cite{Pellegrini2016}, provide a fundamentally different mechanism for high-power ultrashort X-ray generation. 
Through the self-amplified spontaneous emission (SASE) process~\cite{kondratenko1980sase,bonifacio1984collective}, XFELs can deliver photon pulses with excellent spectral tunability, microjoule- to millijoule-level pulse energy, and ultrashort durations down to the attosecond scale.
Such X-ray pulse durations are typically achieved either by generating an ultrashort electron bunch through nonlinear compression at a reduced charge~\cite{huang2017generating,malyzhenkov2020single,prat2023coherent,prat2025enhanced}, or by restricting the X-ray lasing process to a small longitudinal subset of a long electron bunch~\cite{marinelli2017experimental,zhang2020experimental,duris2020tunable,li2024beam,yan2024terawatt,robles2026hardXray}. 

Beyond achieving powerful X-ray pulses with attosecond-scale duration, flexible control over pulse duration, pulse pair configuration, and pulse energy is becoming increasingly important for attosecond XFEL experiments. 
Many ultrafast phenomena unfold on femtosecond timescales comparable to the Auger-Meitner (AM) lifetime~\cite{drescher2002time,haynes2021clocking}, which characterizes the decay of inner shell vacancy through electron rearrangement. 
For example, in single-particle imaging experiments, subfemtosecond X-ray pulses outrun radiation damage~\cite{kuschel2025non}, while tens-of-femtosecond X-ray pulses exceed the timescale over which ionic structural damage happens through AM decay. 
A similar consideration arises in serial femtosecond crystallography~\cite{Chapman2011Femtosecond}, where electronic damage can develop on few-femtosecond to attosecond timescales~\cite{Nass2015Indication}. 
The ability to vary X-ray pulse durations from tens-of-femtosecond to attosecond provides direct insight into ultrafast dynamics driven by electron-electron correlation. 
Two-colour X-ray pump-probe pulse pairs provide a complementary route to resolving ultrafast dynamics with elemental specificity and time resolution down to the attosecond level. 
By tuning the pump and probe photon energies to different core-level resonances, an electronic excitation can be initiated at one atomic site and interrogated at another~\cite{lutman2013experimenal,al2022observation,Prat2022}. 
As the pump pulse duration approaches the attosecond scale, the interaction can approach the impulsive limit of electronic motion, opening opportunities for X-ray Raman excitation of coherent neutral electronic wavepackets~\cite{o2020electronic}, whose subsequent evolution can be tracked with a delayed probe pulse.
Meanwhile, advanced nonlinear X-ray spectroscopy techniques involving multiple-photon processes such as X-ray four-wave-mixing~\cite{Morillo-Candas2026Coherent} benefit from both attosecond temporal resolution and high pulse intensities. 
Maximizing the pulse energy of isolated attosecond X-ray pulses is thus an important target of FEL source development efforts.

The photocathode laser temporal shaping technique~\cite{zhang2020experimental}, first demonstrated at the soft X-ray beamline of the Linac Coherent Light Source (LCLS), generates a short high-current spike at the core of a long electron bunch, suitable for producing an isolated attosecond FEL pulse.
Furthermore, the coexistence of the spike and pedestal enables a broad range of advanced FEL operation modes such as generation of phase-locked harmonic pump-probe pulse pairs~\cite{guo2024experimental}, terawatt-scale attosecond X-ray pulses via multistage amplification~\cite{lutman2018high,Wang2024Fresh,franz2024terawatt}, and spectrotemporal shaping through frequency pulling~\cite{robles2025spectrotemporal}.

This work features the SwissFEL soft X-ray undulator beamline Athos (Methods~\ref{sec:Methods_SwissFEL}), which provides additional degrees of flexibility through its unique layout~\cite{prat2020compact, prat2023x}. 
Unlike conventional XFEL beamlines consisting of mainly planar undulators and a limited number of variable-polarization afterburner modules~\cite{schneidmiller2013obtaining,lutman2016polarization}, Athos consists entirely of variable-polarization \mbox{APPLE-X} undulator modules, which can continuously transition between planar and helical configurations~\cite{schmidt2018apple,kittel2024demonstration}.
Athos is thus uniquely suitable for generating the bright, fully polarization-controlled attosecond X-ray pulses that so far remain inaccessible to table-top sources. 
Helical undulators also provide intrinsic advantages for attosecond XFEL operation, since the radiation field couples to the electron beam in two orthogonal transverse planes, effectively resulting in a stronger FEL interaction than in the case of planar undulators. 
This increased coupling leads to a higher FEL gain, a shorter gain and saturation length, and a reduced cooperation (coherence) length compared to linearly polarized operation at the same photon energy and electron beam configuration~\cite{Kittel2024Helical}. 
As a result, circularly polarized attosecond XFEL pulses can develop broader single-spike spectral bandwidths and shorter temporal durations than their linearly polarized counterparts. 
Such intrinsically broader coherent bandwidths are particularly advantageous for advanced XFEL schemes such as impulsive stimulated X-ray Raman scattering~\cite{o2020electronic}.
Additionally, Athos is equipped with small delaying chicanes placed between neighboring undulator modules, which provide a high degree of flexibility for setting up multistage amplification schemes.
The low-charge nonlinear compression scheme has previously been demonstrated at Athos, providing isolated attosecond-scale X-ray pulses with variable polarization and average pulse energies of up to 20~\textmu J~\cite{prat2023coherent,Prat2025Polarix}.

In this article, we report the generation of high-power attosecond soft X-ray pulses across the 450--1070~eV photon energy range and with either linear or circular polarization, together with the capability for rapid switching between attosecond and few-femtosecond pulse-duration regimes. 
We show that our electron beam in this configuration exhibits a time-dependent transverse orbit, which allows us to selectively confine the XFEL interaction to different temporal regions of the bunch.
Thus, the current spike can generate isolated attosecond pulses, while the pedestal can support longer femtosecond pulses. 
Furthermore, we realize two-colour X-ray pump-probe operation with independently tunable photon energies and delay, as well as multistage amplification of circularly polarized attosecond-scale pulses. 
Together, these capabilities establish a powerful platform for polarization-controlled and pulse duration-dependent nonlinear attosecond soft X-ray science.

\section{Results}\label{Sec:Results}

\subsection{Electron-beam phase-space shaping}\label{sec:Experiment_Setup}

\begin{figure*}
\centering
\includegraphics[width=\linewidth]{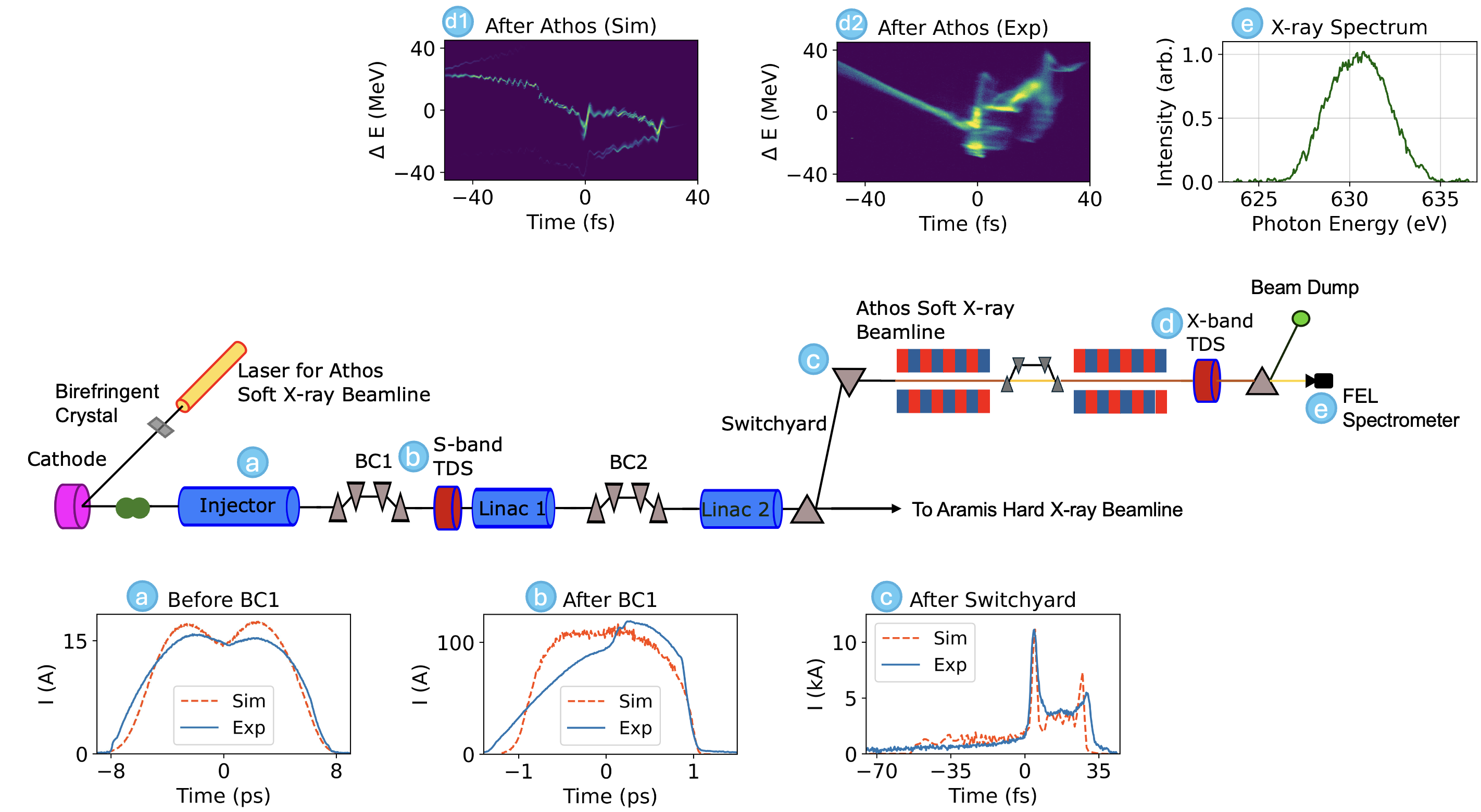}
\caption{\textbf{Facility layout and key results.} 
(a-e) Measured (Exp) and simulated (Sim) longitudinal phase-space (LPS) and current profiles of the shaped electron beam through the accelerator, with the bunch head on the right side.
(a)~Before BC1 compression, showing a local current minimum in the bunch core.
(b)~After BC1 compression, the local current minimum disappears, indicating a stronger bunch compression at the core. 
(c)~After the switchyard, showing a middle current spike with approximately 10~kA peak current and a FWHM\ duration below 5~fs. 
(d1,d2)~Simulated and measured LPS profiles in the X-band TDS at the Athos soft X-ray beamline exit. 
Details on LPS and current profile measurements are discussed in Methods~\ref{sec:Methods_Diagnostics}. 
(e)~Measured single-spike soft X-ray spectrum.
}
\label{fig:Exp_Schematic_and_Beam_Dynamics}
\end{figure*}

Figure~\ref{fig:Exp_Schematic_and_Beam_Dynamics} summarizes the experimental measurements and simulations of electron beams generated with the photocathode laser temporal shaping mode for attosecond-scale soft X-ray pulse generation at SwissFEL (Methods~\ref{sec:Methods_Diagnostics} and \ref{sec:Methods_S2E_Simulation}).
The double-Gaussian current profile before the first bunch compressor~(BC1), shown in Fig.~\ref{fig:Exp_Schematic_and_Beam_Dynamics}a, is achieved by shaping the photocathode laser pulse with a birefringent crystal (Methods~\ref{sec:Methods_Photocathode_Shaping_Setup}). 
This initial density modulation gives rise to longitudinal space charge (LSC) forces within the injector section, which introduce a nonlinear energy chirp that is qualitatively preserved until the end of the second section of the SwissFEL main linear accelerator~(linac 2), even though the current profile significantly changes at each bunch compression stage (see Supplementary Information Sec.~\ref{sec_SI:residual_nonlinear_chirp}). 
Since high-energy particles take shorter paths than low-energy particles through all three bunch-compression sections, the nonlinear chirp at the bunch core enhances the local compression factor.
BC1 removes the initial current minimum (Fig.~\ref{fig:Exp_Schematic_and_Beam_Dynamics}b).
The combination of the second bunch compressor (BC2) and the Athos extraction line (also called switchyard) produces a central current spike with a peak of around 10~kA and a width below 5~fs (Fig.~\ref{fig:Exp_Schematic_and_Beam_Dynamics}c), while the entire bunch is about 100~fs long.

Within the two bunch compressors and the switchyard, the current spike formed in the bunch core experiences stronger coherent synchrotron radiation~(CSR~\cite{saldin1997coherent}) kicks, resulting in a transverse orbit that differs from the surrounding pedestal current. 
In addition, a residual transverse lattice dispersion after the switchyard may further introduce energy-dependent transverse orbit variations within the electron beam. 
This CSR-induced transverse phase-space separation is the key mechanism underlying the generation of isolated, attosecond-scale soft X-ray pulses, and also underpins three advanced machine operation modes demonstrated later in this work, where the lasing slice can be selected via orbit control. 
Additionally, the large current spike generated in the switchyard naturally develops a strong internal energy chirp through LSC forces, which act in opposite direction to those in the injector. 
Longitudinal phase-space (LPS) measurements performed with the X-band transverse deflecting structure (TDS) downstream of the Athos beamline show that the peak-to-peak energy modulation within the central current spike reaches several tens of MeV (Fig.~\ref{fig:Exp_Schematic_and_Beam_Dynamics}d1,d2). 
This large energy chirp further supports the generation of attosecond-scale X-ray pulses from the spike and the suppression of lasing from the pedestal current via the chirp-taper matching scheme~\cite{Saldin2006}.

\subsection{Measurement and modeling of attosecond-scale soft X-ray pulses}\label{sec:measurements_and_modeling}

\begin{figure*}
\centering
\includegraphics[width=\linewidth]{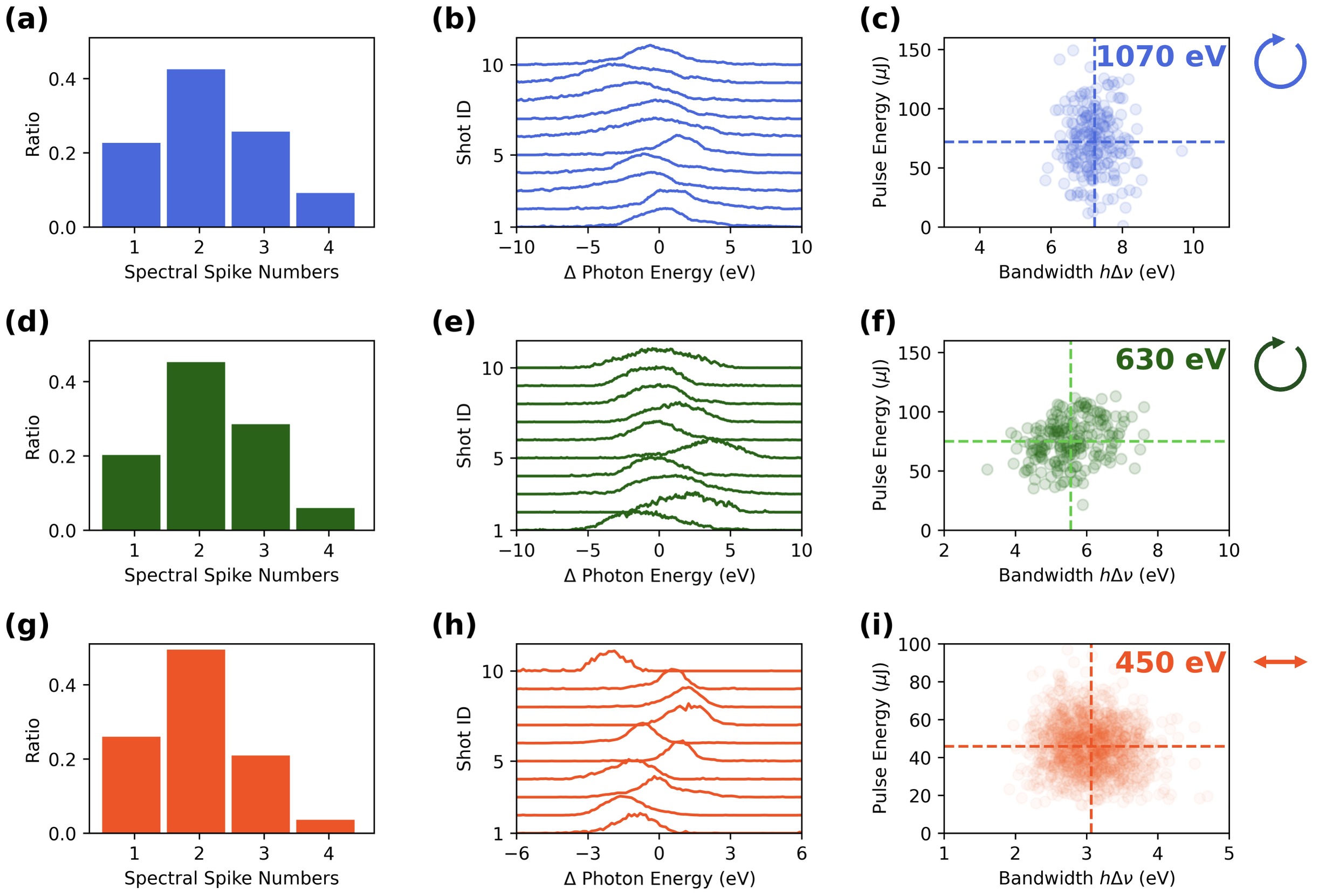}
\caption{\textbf{Analysis of measured X-ray pulses.} 
Spectral peak-counting statistics, example single-spike spectra, and pulse-energy–bandwidth correlations for photon energies of 1070~eV~(a–c), 630~eV~(d–f), and 450~eV~(g–i). 
The X-ray pulses have circular polarization at 1070~eV and 630~eV, and linear polarization at 450~eV. 
(a,d,g)~Histograms of spectral-spike numbers. 
(b,e,h)~Ten representative single-shot spectral measurements containing only one spectral spike.
(c,f,i)~Scatter plots of spectral bandwidths and pulse energies for single-spike shots. 
Horizontal and vertical dashed lines in panels (c, f, i) indicate the average pulse energy and bandwidth, respectively, for each configuration. 
Data in this figure are based on analyses of 1000, 1000, and 5000 consecutive shots for the 1070~eV, 630~eV, and 450~eV cases, respectively.
}
\label{fig:Exp_Data_on_XFEL_Pulses}
\end{figure*}

After aligning the transverse orbit and matching the undulator taper to the energy chirp of the current spike, LPS measurements with the X-band TDS downstream of Athos confirm that only the spike is lasing. 
This observation provides the first indication that the X-ray pulse duration is ultrashort. 
However, the duration cannot be directly determined from the LPS measurement with subfemtosecond resolution. 
This limitation arises for two reasons: first, the temporal resolution of the X-band TDS in our configuration is approximately 2~fs root-mean-square (rms) (see Methods~\ref{sec:Methods_Diagnostics}); and second, due to radiation slippage in the undulator, the temporal length of the energy loss region within the current spike is longer than the X-ray pulse duration. 
The slippage effect is generally more significant at lower X-ray photon energy.

To further characterize the ultrashort X-ray pulses, we analyze the single-shot photon spectra exhibiting a single dominant spike with multi-eV bandwidth (Fig.~\ref{fig:Exp_Schematic_and_Beam_Dynamics}e). 
Subfemtosecond X-ray pulses intrinsically have broad, coherent bandwidths. 
Consequently, isolated X-ray pulses with single-spike spectra and multi-eV bandwidths serve as strong experimental indications of attosecond-scale emission.
For instance, the full-width-at-half-maximum (FWHM) duration $\Delta t_{\mathrm{FWHM}}$ and the FWHM frequency bandwidth $h\Delta \nu_{\mathrm{FWHM}}$ of a time-bandwidth limited Gaussian pulse fulfill the condition
\begin{equation}
h\Delta \nu_{\mathrm{FWHM}} \cdot \Delta t_{\mathrm{FWHM}} = 1.83~\mathrm{eV}\cdot\mathrm{fs}.
\label{eq:ftl}
\end{equation}
In practice, subfemtosecond X-ray pulses are not ideal Gaussian waveforms; instead, they exhibit more complex spectral intensities and phase chirps. 
As a result, estimating the pulse duration via Eq.~(\ref{eq:ftl}) generally only provides a lower bound on the actual temporal pulse length. 
Nevertheless, angular streaking measurements at LCLS indicate that the average pulse duration in comparable XFEL operating modes remains within approximately a factor of two of the Fourier transform limit~\cite{duris2020tunable}.

The statistical properties of the emitted XFEL pulses are summarized in Fig.~\ref{fig:Exp_Data_on_XFEL_Pulses}. 
The measurements cover three photon energies: 1070~eV and 630~eV in circular polarization, and 450~eV in linear polarization. 
Panels (a,d,g) show the corresponding distributions of spectral-spike numbers (Methods~\ref{sec_SI:spectral_peak_counting}). 
At each X-ray photon energy, more than 50\% of the shots contain only one or two spectral spikes, confirming FEL operation in the ultrashort pulse regime. 
Panels (b,e,h) show representative single-shot spectra with only one spectral peak and exhibiting a multi-eV bandwidth, consistent with attosecond-scale pulse durations.
Panels (c,f,i) display scatter plots of the measured X-ray pulse energies and spectral bandwidths for single-spike shots (Methods~\ref{sec:Methods_bandwidth_pulse_length_determination}), showing no significant correlation. 
The average bandwidths and pulse energies of single-spike shots reach 7.2~eV and 70~\textmu J at 1070~eV, 5.6~eV and 70~\textmu J at 630~eV, and 3.1~eV and 40~\textmu J at 450~eV. 

\begin{figure*}
\centering
\includegraphics[width=\linewidth]{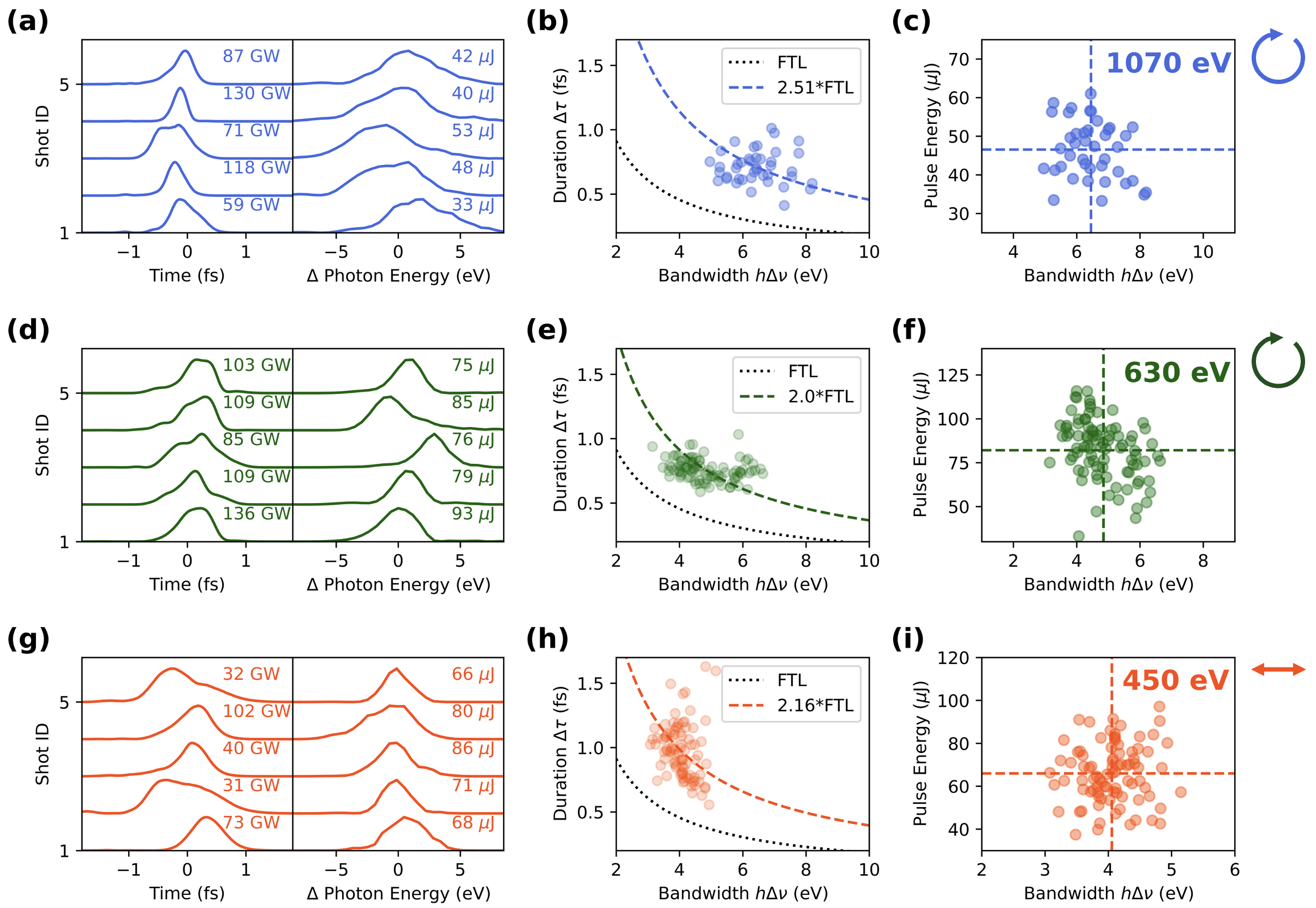}
\caption{\textbf{Analysis of simulated X-ray pulses.} 
Simulated XFEL pulses with single-spike spectra at photon energies of 1070~eV~(a–c), 630~eV~(d–f), and 450~eV~(g–i). 
Panels~(a,d,g) show representative temporal profiles and corresponding spectra for five shots, normalized by the peak intensities. 
The simulated peak power and pulse energy are reported for each pulse profile.
Panels~(b,e,h) show pulse duration $\Delta\tau$ versus spectral bandwidth $h\Delta\nu$, with the Fourier transform limit~(FTL, dotted) and the average $\Delta\tau \cdot h\Delta\nu$~(dashed). 
Panels (c,f,i) show pulse energies versus spectral bandwidths. 
Horizontal and vertical lines indicate the average pulse energy and bandwidth, respectively. 
Each simulation dataset is based on 250 simulated shots with different random seeds.
}
\label{fig:Simulations}
\end{figure*}

To support and interpret the experimental observations, we reproduced the X-ray spectral measurements in start-to-end simulations (Methods~\ref{sec:Methods_S2E_Simulation}). 
Figure~\ref{fig:Simulations} shows simulated single-spike X-ray pulses in both the temporal and spectral domains, along with correlations between spectral bandwidth and pulse duration.
Panels (a--c), (d--f), and (g--i) correspond to photon energies of 1070~eV (circular polarization), 630~eV (circular polarization), and 450~eV (linear polarization), respectively. 
The percentage of simulated shots containing only one spectral spike (two spectral spikes) is approximately 20\%--35\% (50\%--60\%), in reasonable agreement with the measurements discussed earlier~(Fig.~\ref{fig:Exp_Data_on_XFEL_Pulses}a,d,g).
Each single-spike pulse reaches multi-eV spectral bandwidth, consistent with the experimental observations~(Fig.~\ref{fig:Simulations}a,d,g).
Across simulated single-spike shots, the average pulse-length-bandwidth products (Methods~\ref{sec:Methods_bandwidth_pulse_length_determination}) are approximately twice the Fourier transform limit (Fig.~\ref{fig:Simulations}b,e,h).

Based on the measured bandwidths and the simulated time-bandwidth correlations, we infer average pulse durations of about 0.6~fs at 1070~eV, 0.8~fs at 630~eV, and 1.2~fs at 450~eV. Combined with the measured pulse energies, this corresponds to average X-ray powers of approximately 100~GW at 1070~eV and 630~eV, and 35~GW at 450~eV. 
The simulations further show no significant correlation between spectral bandwidths and pulse energies for single-spike shots~(Fig.~\ref{fig:Simulations}c,f,i), in agreement with the experimental measurements.

\subsection{X-ray pulse length control via transverse orbit}\label{sec:XFEL_Simulation}

\begin{figure*}
\centering
\includegraphics[width=\linewidth]{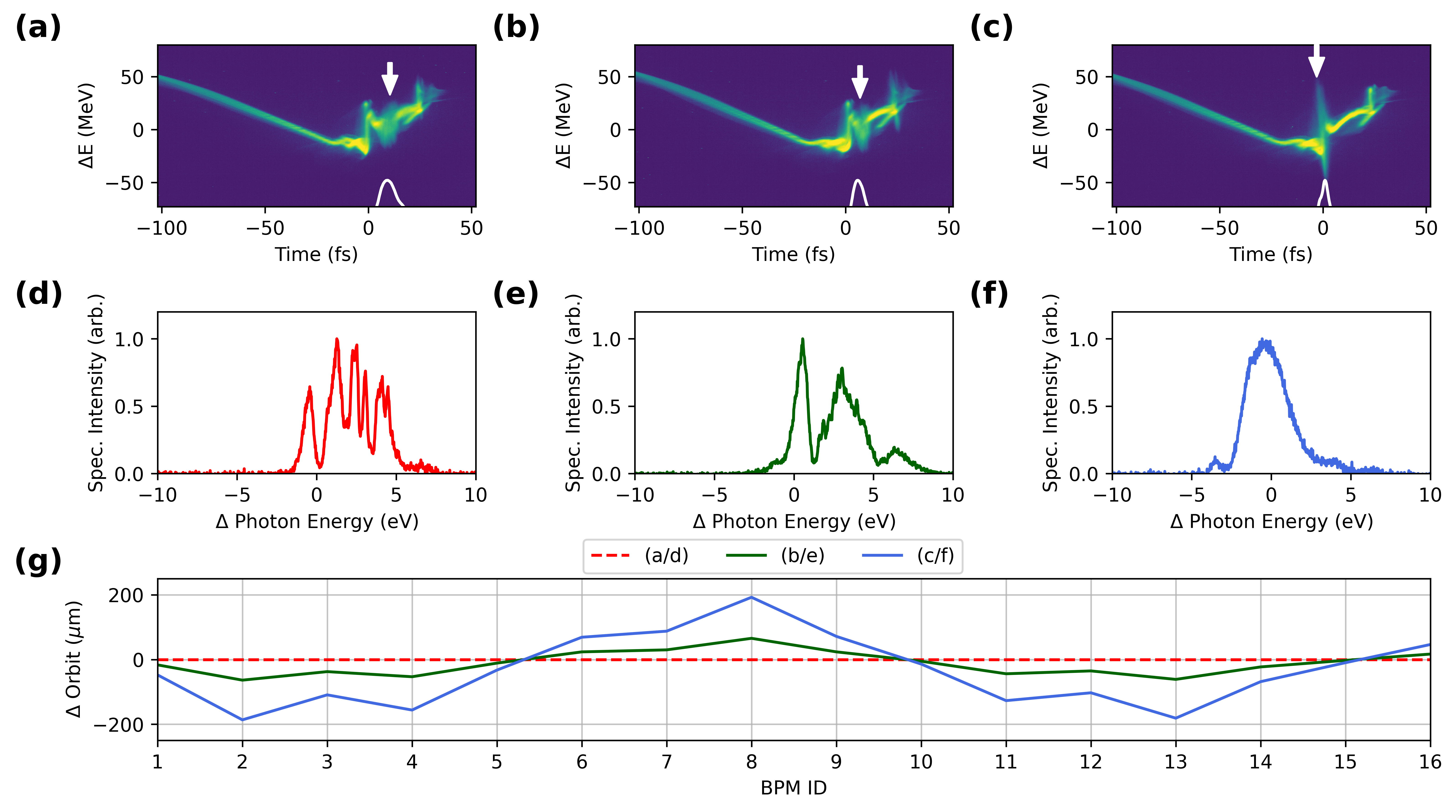}
\caption{\textbf{Pulse length control via transverse orbit.} 
Measured LPS profiles of electron beams (a–c) with the bunch head on the right side and corresponding X-ray spectra (d–f) for three lasing conditions at 630~eV photon energy: FWHM\ 7~fs~(a,d), FWHM\ 4~fs~(b,e), and FWHM\ subfemtosecond~(c,f) X-ray pulses. 
White arrows in panels (a–c) mark the lasing slice within the electron bunch. 
White lines at the bottom of panels (a-c) show the longest limits of XFEL temporal pulse profiles reconstructed from the electron beam LPS measurements. 
Panel (g) shows the center-of-mass horizontal orbits measured with the beam-position monitors in the Athos soft-X-ray undulator beamline, relative to the orbit for the 7~fs configuration. 
}
\label{fig:Pulse_Length_Control}
\end{figure*}

To demonstrate flexible X-ray pulse duration control, we exploit the distinct transverse orbits of different temporal slices of the electron bunch. 
Through corrector magnet strength adjustments, we can move the X-ray lasing from the front part of the long pedestal to the high-current spike. 
Figure~\ref{fig:Pulse_Length_Control} (a--f) shows measured LPS profiles and corresponding X-ray spectra for three representative orbit configurations at a central photon energy of 630~eV. 
When lasing occurs on the pedestal and the pulse duration is several femtoseconds long, the spectra of emitted X-ray pulses contain multiple narrow SASE spikes (Fig.~\ref{fig:Pulse_Length_Control}a,d), reflecting the presence of many longitudinal modes. 
Through adjustment of a single corrector dipole before the undulator section, the lasing region is shifted towards the core and slightly reduced in duration (Fig.~\ref{fig:Pulse_Length_Control}b,e). 
A further corrector dipole adjustment confines the lasing region to the high-current spike, generating an ultrashort X-ray pulse as indicated through single-spike spectra with multi-eV bandwidths (Fig.~\ref{fig:Pulse_Length_Control}c,f). 
The corresponding measured horizontal beam centroid trajectories along the Athos undulator line are shown in Fig.~\ref{fig:Pulse_Length_Control}g. 
The measured difference in horizontal centroid orbit under different lasing conditions is in good agreement with simulations (Supplementary Information Sec.~\ref{sec_SI:transverse_phase_space}). 
From the LPS measurements we estimate pulse durations of around 7 and 4~fs for the first two cases, respectively. 
Based on the measured X-ray bandwidth in Fig.~\ref{fig:Pulse_Length_Control}f and the benchmarking simulation, we estimate the X-ray pulse duration to be around 700~as in the third and shortest case. 
We emphasize that the pulse-duration switching demonstrated here was done via a field strength adjustment of a single corrector dipole.
This requires only a few seconds, although several other machine parameters would typically be empirically tuned in order to optimize the FEL performance whenever the working point is changed.

\subsection{Two-Colour X-ray Pulse Pair Generation}\label{sec:Two_Color}

\begin{figure}
\centering
\includegraphics[width=\linewidth]{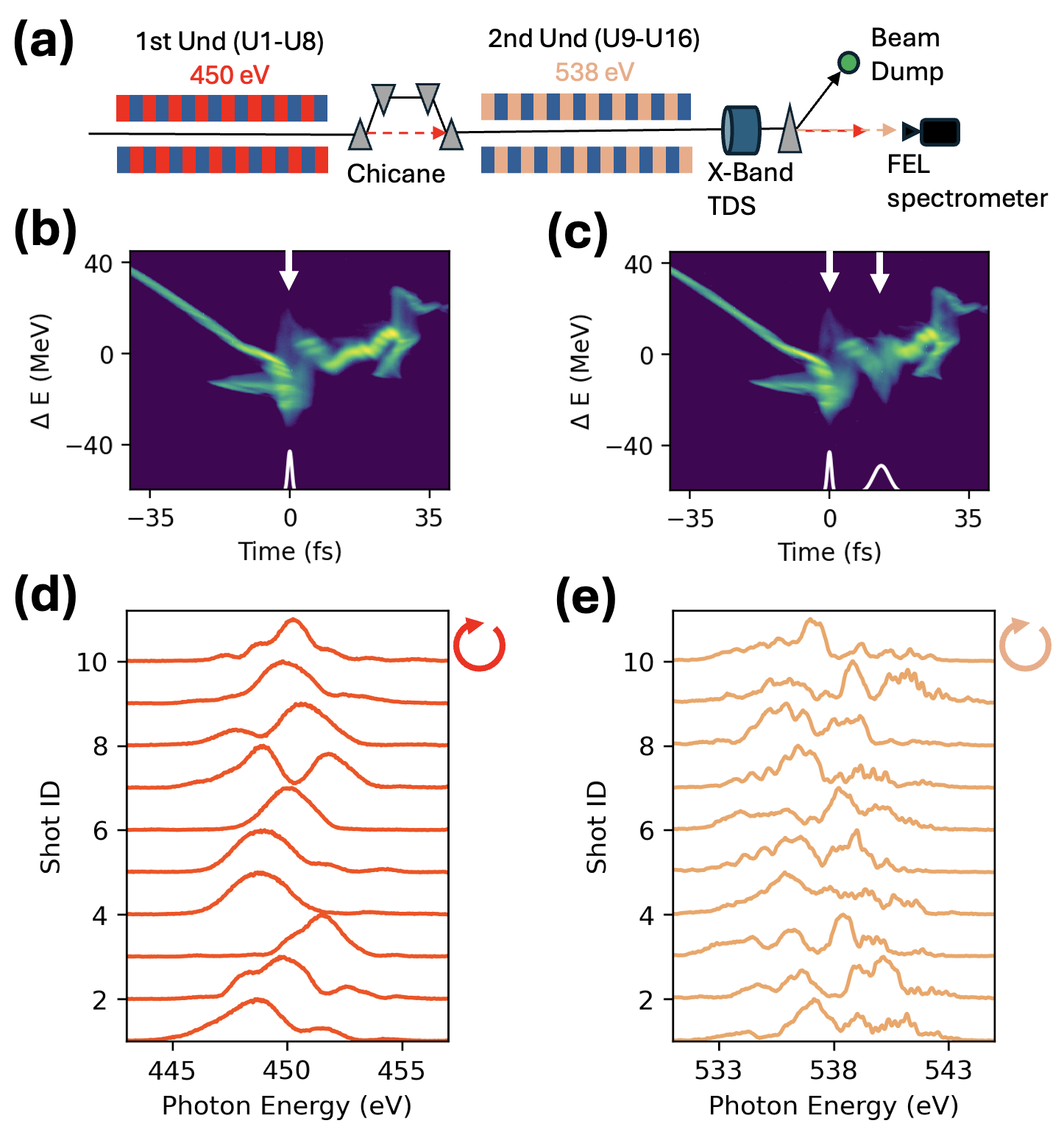}
\caption{\textbf{Generation of two-colour X-ray pulse pairs.} 
(a) Athos undulator configuration used for two-colour operation. 
(b,c) Measured LPS profiles for lasing from the current spike only (b), and from both the current spike and pedestal (c). 
White arrows mark the lasing regions. 
White traces indicate the corresponding X-ray pulse structures. 
(d,e) Ten consecutive single-shot spectral measurements of the 450 eV attosecond pulse and the 540 eV few-femtosecond pulse in the two-colour operation.
}
\label{fig:Two_Color}
\end{figure}

The ability to select different temporal slices of the electron beam for lasing can also be exploited to generate two-colour X-ray pulse pairs with independently tunable photon energies and controllable delay in the split-undulator configuration~\cite{lutman2013experimenal,Prat2022}. 
Figure~\ref{fig:Two_Color}a shows the Athos undulator set-up used for this operation mode. 
The undulator line is divided into two sections separated by a two-colour chicane, which can delay the electron bunch with respect to the X-ray pulse by up to about 500~fs. 
The current spike is aligned to the lasing orbit in the first undulator section and generates a circularly polarized attosecond X-ray pulse at 450~eV, while the pedestal current is aligned to the lasing orbit in the second section to generate a circularly polarized few-femtosecond pulse at 540~eV. 
By independently tuning the undulator fields in the two sections, the photon energies of the pump and probe pulses can be freely adjusted. 
The two-colour chicane provides a continuous control of the pump–probe delay, allowing the relative arrival time of the two pulses to be tuned through the zero crossing. 

The lasing process is directly visible in the measured LPS profiles shown in Fig.~\ref{fig:Two_Color}b,c. 
When only the current spike lases, energy loss is observed exclusively from within the central current spike (Fig.~\ref{fig:Two_Color}b). 
When both undulator sections are resonant, an additional energy loss region appears in the pedestal current, demonstrating generation of the second X-ray pulse (Fig.~\ref{fig:Two_Color}c). 
The average X-ray spectra remain nearly unchanged when each colour is operated individually or when both colors lase simultaneously (Supplementary Information Sec.~\ref{sec_SI:two_color_average}), indicating negligible mutual influence between the two lasing processes.
The average pulse energies are 52~\textmu J at 450~eV and 62~\textmu J at 540~eV. 
The average bandwidth of single-spike spectra recorded at 450~eV is 3.0~eV, corresponding to an estimated pulse duration of 1.2~fs when assuming a time-bandwidth product of twice the Fourier transform limit, which is below the resolution limit of LPS measurement. 
The FWHM\ duration of the 540~eV pulse is estimated at below 5~fs from the measured LPS distribution.
Representative single-shot spectra of both pulses are shown in Fig.~\ref{fig:Two_Color}d,e. 

\subsection{Multistage Amplification of Attosecond X-ray Pulse}\label{sec:Attosecond_MSA}

\begin{figure*}
\centering
\includegraphics[width=\linewidth]{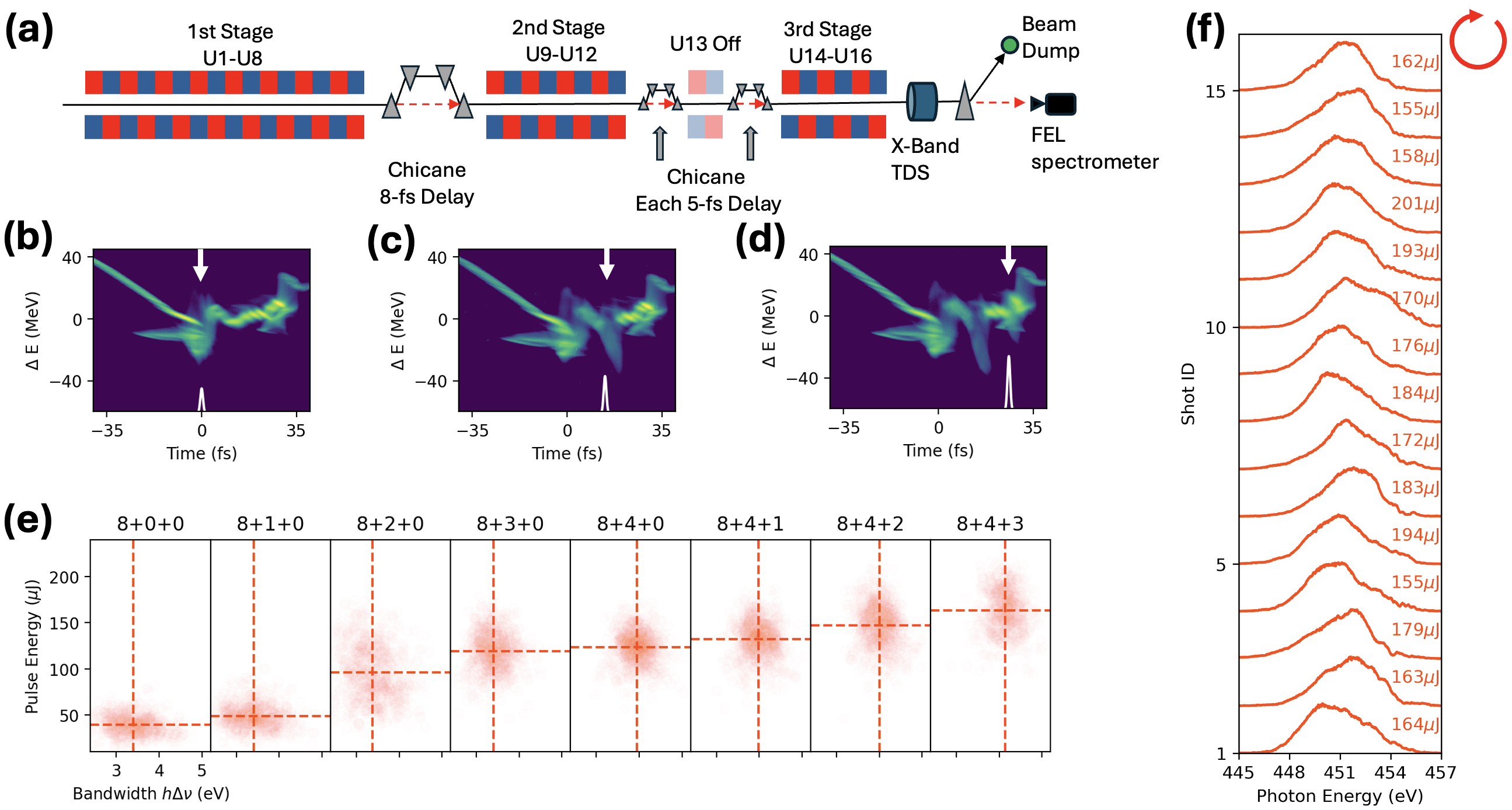}
\caption{\textbf{Three-stage amplification of attosecond X-ray pulse.} 
(a) Athos undulator configuration used for three-stage amplification of attosecond X-ray pulses.
(b,c,d) Measured electron-beam LPS for one-stage, two-stage, and three-stage amplification. 
White arrows indicate the lasing slices, while the white traces schematically represent the corresponding X-ray pulse. 
(e) Spectral bandwidth and pulse energy for single-spike shots as the number of undulator modules for amplification is increased. 
The panel titles denote the number of undulator modules used in each amplification stage. 
Dashed lines indicate average values. 
(f) Representative single-shot spectra and pulse energies after three-stage amplification.
}
\label{fig:Attosecond_MSA}
\end{figure*}

We exploit the same slice-selective lasing capability to increase the pulse energy beyond the saturation power of the high-current spike through multistage amplification (MSA), where the attosecond pulse generated by the current spike in the first half of the undulator is overlapped with fresh electrons in the pedestal to achieve additional amplification in the second half of the undulator. 
Figure~\ref{fig:Attosecond_MSA}a shows the Athos undulator configuration used for the MSA experiment at 450~eV with circular polarization. 
Delaying chicanes divide the undulator line into three stages consisting of 8, 4, and 3 undulator modules. 
An isolated attosecond pulse is first generated by the current spike in the first stage (Fig.~\ref{fig:Attosecond_MSA}b). 
The two-colour chicane then delays the electron beam by approximately 8~fs relative to the X-ray pulse.
We then configure the orbit control system to restore lasing conditions to the pedestal part for the next undulator modules, leading to further amplification (Fig.~\ref{fig:Attosecond_MSA}c).
When more than four modules are used, additional lasing regions corresponding to an unwanted SASE background become visible in the LPS.
We suppress the SASE background with an additional amplification stage using further chicanes that wash out the accumulated microbunching in the pedestal current.
We find that a single small chicane does not provide enough longitudinal dispersion, therefore we realize the second delaying stage with a combination of two small chicanes and an off-resonant undulator module, which in total provide a further delay of 10.4~fs (Fig.~\ref{fig:Attosecond_MSA}d). 

The evolution of pulse energy and spectral bandwidth on shots with a single spectral spike during the amplification process is summarized in Fig.~\ref{fig:Attosecond_MSA}e. 
The ratio of shots with single-spectral spikes is above 60\% across the entire MSA process. 
At the end of the first stage (``8+0+0"), the average pulse energy and bandwidth are 40~\textmu J and 3.4~eV, respectively. 
The second stage (``8+4+0") increases these values to 123~\textmu J and 4.0~eV, and the third stage (``8+4+3") increases them to 163~\textmu J and 4.1~eV.
The highest measured single-shot pulse energy and bandwidth are 217~\textmu J and 5.0~eV, respectively. 
Representative single-shot spectra after the third stage are shown in Fig.~\ref{fig:Attosecond_MSA}f, while spectral peak-counting statistics are included in Supplementary Information Sec.~\ref{sec_SI:More_on_MSA}.

In summary, MSA increases the average pulse energy by approximately a factor of four while preserving the broad single-spike spectral structure.
The continuous increase in bandwidth during re-amplification is consistent with the soliton-like superradiant regime reported in Ref.~\cite{franz2024terawatt}. 
Assuming a time-bandwidth product of twice the Fourier transform limit, the average pulse duration is approximately 0.9~fs and the average peak power is about 185~GW.

\section{Conclusion}\label{sec:Conclusion}

We have generated high-power, attosecond-scale soft X-ray pulses across a broad photon energy range (450--1070~eV) and with either linear or circular polarizations at SwissFEL. 
This capability is enabled by a combination of the photocathode laser temporal shaping technique and the full polarization tunability of the \mbox{APPLE-X} undulator modules in the Athos beamline. 
A single birefringent crystal introduced into the photocathode laser produces a stable double-Gaussian initial current profile, which evolves into an isolated high-current spike on top of a low-current pedestal. 
This current spike drives X-ray emission with average pulse energies of 40--70~\textmu J and estimated pulse durations in the range of 0.6--1.2~fs.
These average pulse energies exceed those that have been achieved with the low-charge mode at SwissFEL Athos~\cite{prat2023coherent,Prat2025Polarix}, and are comparable to those that were achieved at the LCLS-I soft X-ray facility~\cite{duris2020tunable,zhang2020experimental}. 

Beyond attosecond pulse generation from the current spike itself, the time-dependent orbit structure of the electron beam supports several advanced operating modes. 
We have demonstrated rapid switching between attosecond and few-femtosecond pulse durations and generated two-colour X-ray pulse pairs with independently tunable photon energies and delay, with one of the pulses having attosecond-scale duration and the other being few femtoseconds long.
Finally, we achieved multistage amplification of circularly polarized attosecond pulses to energies exceeding 200~\textmu J, which is again comparable to LCLS-I results with linear polarization~\cite{franz2024terawatt}. 

The natural next step for this research program is to exploit the versatile and high-power X-ray pulses in end-station experiments that probe ultrafast charge, spin, and chiral dynamics with attosecond temporal resolution and element specificity. 
The photocathode laser shaping concept will also be applied at SwissFEL's second undulator beamline~\cite{prat2020compact} to generate high-power isolated hard X-ray pulses.
The pulse energy achieved in the two-colour and MSA schemes is currently limited by the available fresh electrons in the pedestal current and by the Athos active undulator length of only 32~m. 
In future work we will therefore explore how the shaped temporal profile of the photocathode laser can be engineered to tailor the electron beam current distribution, in particular to optimize the pedestal current available for advanced attosecond XFEL operation modes.
Additional performance improvements are also expected after the proposed SwissFEL\textit{plus} facility upgrade~\cite{Bostedt2026SwissFELPlus}, which includes additional undulator modules for Athos.

\section{Methods}\label{Sec:Methods}

\subsection{SwissFEL}\label{sec:Methods_SwissFEL}

SwissFEL delivers high-brightness ultrashort X-ray pulses from two independently tunable beamlines, Athos for soft X-ray and Aramis for hard X-ray, covering photon energies from 0.25 to 12.4~keV~\cite{prat2020compact,prat2023x}.
Simultaneous operation of both beamlines at a repetition rate of up to 100~Hz is achieved by two independent photocathode lasers, which generate electron bunch pairs with a time separation of 28~ns, and a radio-frequency (RF) system that enables fine-tuning of the RF field seen by the second bunch~\cite{Paraliev2022}.
The accelerator configuration for the second beam towards the Athos beamline contains three bunch-compression stages~(Fig.~\ref{fig:Exp_Schematic_and_Beam_Dynamics}): two four-dipole chicanes (BC1 and BC2) and a dogleg (switchyard).
At a beam energy of 3.4~GeV, the second bunch enters the soft X-ray undulator beamline Athos, which consists of 16~\mbox{APPLE-X} undulators, allowing the generation of arbitrary X-ray polarization~\cite{schmidt2018apple}.
Each undulator module has 52 periods with a period length of 38~mm.

\subsection{Photocathode laser modifications}
\label{sec:Methods_Photocathode_Shaping_Setup}

For the initial experimental studies, we employed both photocathode laser systems, normally dedicated to parallel soft and hard X-ray operation, to form a single bunch with double-Gaussian temporal profile and freely tunable peak separation. 
Once the optimal separation of about 5~ps was determined, a 6.25-mm alpha-BBO birefringent crystal was introduced at one of the laser systems, enabling it to generate the required shape in compatibility with SwissFEL's double bunch operation scheme~\cite{Paraliev2022}.
We find that the available tuning knobs, in particular the gun RF phase and the bunch charge, are sufficient for operating the scheme with the fixed split-and-delay stage.
All results shown here were acquired using just one laser system and with the hard X-ray beamline in standard operation.

\subsection{Accelerator setup}\label{sec:Methods_Accelerator_Setup}

Optimization of the accelerator configuration was performed across multiple accelerator parameters, including the injector S-band RF phase, the X-band linearizer RF voltage, and the RF phase of the first segment of the SwissFEL main linac (linac 1). 
These parameters were tuned empirically to achieve the desired LPS structure (i.e., a single current spike at the center of a longer bunch).

The transverse orbit of the current spike differs substantially from that of the surrounding pedestal current due to CSR effects. 
Several orbit correctors were used to align the trajectory of the high-current spike within the undulator beamline. 
The optimal orbit was identified experimentally by monitoring the X-ray spectra: a high ratio of single-spike spectra indicated correct alignment of the current spike.

Following orbit alignment, a final optimization loop was performed across all relevant accelerator and undulator parameters to maximize the measured X-ray pulse energy and the fraction of single-spike spectra. 
These parameters included accelerator RF phases and voltages, orbit corrector strengths, matching-quadrupole settings, the number of active undulator modules, as well as the undulator taper and phase shifters. 

For both the two-colour and MSA operation modes, the current spike was first optimized to lase in the undulator section upstream of the two-colour chicane, using all available modules. 
The orbit correctors downstream of the chicane were then adjusted to selectively align different temporal slices of the pedestal current to the lasing orbit. 
To shorten the X-ray pulse generated from the pedestal, we increased the beam tilt generated in the switchyard from leaked dispersion.
Such a beam tilt is the product of residual beam energy chirp and lattice dispersion, which can be adjusted via the RF settings of linac~1 and quadrupole magnets in the switchyard~\cite{Wang2024Fresh}.

\subsection{Electron beam and X-ray diagnostics}\label{sec:Methods_Diagnostics}

LPS and current-profile measurements described in this work were performed using an S-band TDS after BC1 and an X-band TDS downstream of the Athos undulator~\cite{Craievich2013,Prat2025Polarix}. 
Each TDS streaks the bunch by correlating transverse momentum with longitudinal position, enabling single-shot temporal diagnostics on a downstream high-resolution screen~\cite{Ischebeck2015,Juranic2023}.
We employ the method described in~\cite{Schmidt2020} to compensate pre-existing beam tilts for the analysis of current profile measurements shown in Fig.~\ref{fig:Exp_Schematic_and_Beam_Dynamics}.
We can effectively extend the locations of current profile measurements in two ways.
First, by disabling the BC1 dipoles we can characterize the uncompressed electron bunch with the S-band TDS after BC1 (Fig.~\ref{fig:Exp_Schematic_and_Beam_Dynamics}a). 
Second, because no further compression occurs after the switchyard, the X-band TDS after Athos can provide an accurate measurement of the electron beam current profile at the switchyard exit (Fig.~\ref{fig:Exp_Schematic_and_Beam_Dynamics}c).

The X-ray lasing process causes both energy loss and energy spread growth of the electron beam. 
By comparing the LPS profiles measured with the X-band TDS for lasing-on and -off configurations, we can measure the XFEL power profile and thus infer an upper limit on the XFEL pulse duration~\cite{Behrens2014}. 
The S-band TDS measurement shown in Fig.~\ref{fig:Exp_Schematic_and_Beam_Dynamics}a exhibits an rms time resolution of approximately 20~fs.
For X-band TDS measurements after Athos we expect significant variations in resolution along the bunch due to the phase-space distortions imposed by the photocathode laser temporal shaping scheme.
The energy-loss region within the pedestal current measured in the second stage of the MSA scheme has an rms width of approximately 1.3~fs, providing an estimate of the time resolution associated to measuring this part of the beam.

The soft X-ray spectra shown in Figs.~\ref{fig:Exp_Schematic_and_Beam_Dynamics}--\ref{fig:Pulse_Length_Control} were measured using a photon spectrometer located at the Maloja endstation~\cite{nordgren1989soft}.
The spectrometer is a Scienta XES 350 model equipped with three gratings (1200, 600, and 300~lines/mm) and a slit width below 100~\textmu m.
For measurements of circularly polarized pulses at 630~eV and 1070~eV, the X-ray spectra were recorded using a multichannel phosphor-plate detector.
For measurements of linearly polarized pulses at 450~eV, a 100-\textmu m-thick Ce-doped YAG screen was employed for spectral imaging.
The two-colour and MSA spectra (Figs.~\ref{fig:Two_Color} and \ref{fig:Attosecond_MSA}) were measured with a different spectrometer on the Athos beamline, equipped with 50 and 150~lines/mm gratings.
The pulses were imaged on another Ce:YAG screen and recorded with a 16-bit pco.edge CMOS camera.
In all configurations the instrumental resolution is considerably finer than the multi-eV single-spike bandwidths reported here.
Single-shot pulse energies were measured using a calibrated gas-based pulse-energy monitor~\cite{Juranic2018}.

\subsection{Peak counting on spectral measurements}\label{sec_SI:spectral_peak_counting}

The number of spectral spikes in each single-shot X-ray spectrum is determined using a peak-finding procedure. 
First, a background signal is subtracted from each spectrum. 
The resulting spectra are then smoothed using a Gaussian filter to suppress high-frequency noise, with standard deviations of 0.3~eV, 0.26~eV, and 0.24~eV for the 1070~eV, 630~eV, and 450~eV datasets, respectively. 
Spectral spikes are subsequently identified as local maxima in the smoothed spectra with a minimum prominence threshold of 0.03 when the peak spectral intensity is normalized to 1. 

\subsection{Calculation of X-ray pulse spectral width and temporal duration}\label{sec:Methods_bandwidth_pulse_length_determination}

Spectral bandwidths $h\Delta\nu$ and temporal pulse lengths $\Delta\tau$ of both the experimental data and numerical simulations presented in Figs.~\ref{fig:Exp_Data_on_XFEL_Pulses} and \ref{fig:Simulations} in the main text are evaluated using the following unified analysis procedure.

Given a nonnegative distribution $f(x)$, its mean value $\overline{x}$ and standard deviation $\sigma_{x}$ are defined as
\begin{equation}
\begin{aligned}
\overline{x} = \frac{\int \mathrm{d}x~x f(x)}{\int \mathrm{d}x~f(x)},\quad
\sigma_{x} = \sqrt{\frac{\int \mathrm{d}x~(x-\overline{x})^2 f(x)}{\int \mathrm{d}x~f(x)}}. \label{eq:RMS}
\end{aligned}
\end{equation}
We define the characteristic width of $f(x)$ as $2\sqrt{2\ln 2}\sigma_{x}$, which corresponds to the FWHM\ of $f(x)$ when $f(x)$ is a Gaussian distribution.

For experimental data, the spectral bandwidth $h\Delta \nu$ is obtained by applying Eq.~\ref{eq:RMS} to the measured single-shot spectral intensity $I(h\nu)$ after thresholding to remove noise.
For numerical simulations, both the spectral bandwidths $h\Delta \nu$ and temporal pulse lengths $\Delta \tau$ are calculated using Eq.~\ref{eq:RMS}, where $f(x)$ is replaced by the simulated far-field spectral or temporal intensity, respectively. 

\subsection{Start-to-end simulation setup}\label{sec:Methods_S2E_Simulation}

To validate the experimental results and gain deeper insight into the single-spike XFEL pulse generation process, we performed comprehensive start-to-end simulations that reproduce the electron-beam dynamics throughout the accelerator and the resulting attosecond X-ray pulse formation in the undulator beamline. 
The simulations employ \mbox{ASTRA}~\cite{FlottmannASTRA} for the low-energy injector region (up to 157~MeV), \mbox{elegant}~\cite{borland2000elegant} for the remainder of the linear accelerator, and Genesis~1.3~\cite{Reiche1999Genesis} for modeling the X-ray pulse generation.
For each beam configuration, 250 Genesis 1.3 runs with different random seeds are performed to account for the stochastic nature of the SASE process.

An important parameter in the simulations is the slice energy spread (SES) of the electron beam before BC1. 
With a given energy chirp before the bunch compressor, a lower SES can generate a shorter and higher current spike. 
Previous studies have identified intrabeam scattering (IBS) as a major contributor to the SES at this location, and for the standard Gaussian-shaped photocathode laser pulse and 200~pC bunch charge, a SES of about 6~keV rms was measured after the BC1 location~\cite{prat2022energy}.
The IBS effect scales with the beam current and inversely with the emittance, and for the photocathode laser shaping configuration we measured a larger emittance and a lower current at the bunch core compared to the standard configuration.
From the scaling laws provided in~\cite{prat2022energy} we expect an SES of about 2.5~keV rms at the bunch core before BC1.
As IBS is not included in the simulation codes, we add SES of 3.5~keV rms to the \mbox{ASTRA} output distribution, with its magnitude chosen to also provide a margin for further energy spread increase after BC1. 

\section*{Acknowledgments}
We thank all the technical groups involved in the operation of SwissFEL. 
Z.G., H.Z. and C.B. acknowledge support from the Swiss National Science Foundation under grant number 197372. 

\section*{Author contributions}
Z.G., C.V., E.P., S.R., C.B., and P.D. conceived the research project. 
C.V., A.D., U.D., M.H., and A.T. modified and set up the photocathode laser system. 
All authors participated in the experimental study and interpretation of results. 
Z.G. and P.D. analyzed the data, performed the simulations, and wrote the initial version of the manuscript. 
All authors contributed to the final version of the manuscript.

\clearpage
\section{Supplementary Information}

\subsection{Additional simulation details}\label{sec_SI:additional_simulation}
\subsubsection{Nonlinear energy chirp in bunch core}\label{sec_SI:residual_nonlinear_chirp}

In the one-dimensional (1D) limit, the LSC field $E(z)$ is approximately
\begin{equation}
E(z) = -\frac{Z_0}{4 \pi \gamma^2} \left( 1+2\log\left( \frac{\gamma \sigma_z}{r_b } \right)\right) \frac{\mathrm{d}I(z)}{\mathrm{d}z},\label{eq:E_z_gradient}
\end{equation}
where $Z_0$ is the impedance of free space, $\gamma$ is the Lorentz factor of the electron beam, $\sigma_z$ is the rms bunch length, $r_b$ is the transverse beam size, and $I(z)$ is the current profile~\cite{venturini2008models}. 
Equation~\ref{eq:E_z_gradient} shows that the LSC field is directly proportional to the longitudinal gradient of the current profile, $dI(z)/dz$. 
Consequently, any nonuniform current distribution generates a local energy chirp within the electron beam. 
This mechanism drives the LPS shaping employed in this work.

Equation~\ref{eq:E_z_gradient} further reveals that for a given electron distribution, $E(z)$ is strongest at low beam energies.
Consequently, the shape of the higher-order energy chirp imprinted in the bunch core before the first bunch compressor (i.e., where the beam energy is lowest) is largely preserved throughout subsequent acceleration up to the end of Linac~2, even though the current profile of the beam undergoes substantial modification during compression in BC1 and BC2.
Only after the switchyard, the current spike is large enough for the space-charge forces to impose a chirp with opposite sign.

\begin{figure*}
\centering
\includegraphics[width=0.75\linewidth]{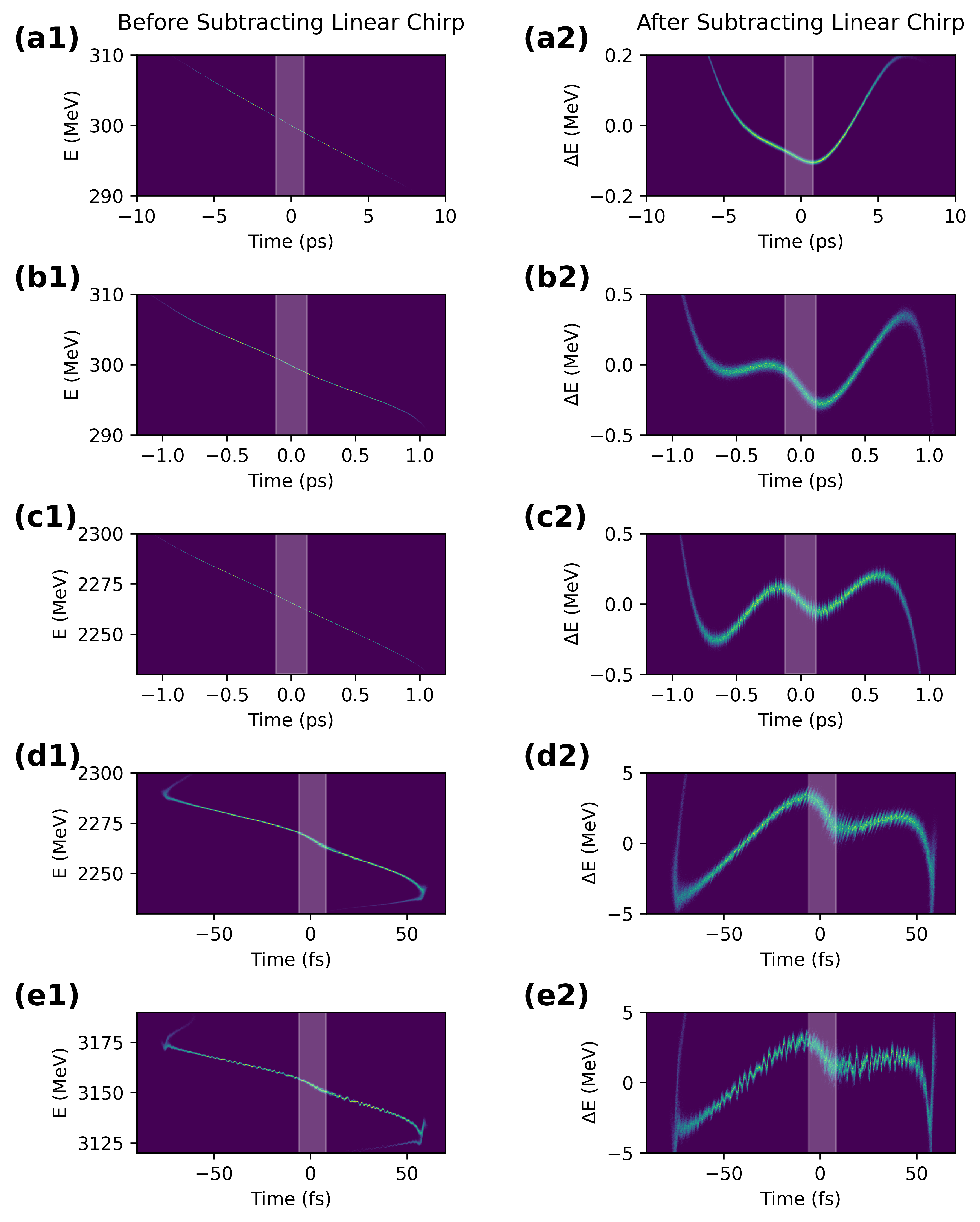}
\caption{\textbf{Residual nonlinear energy chirp in LPS.} 
(a1,a2) Simulated longitudinal phase-space (LPS) of the shaped electron beam before BC1, shown before (a) and after (b) subtracting a linear energy chirp.
(b1,b2) Same analysis at the end of BC1. 
(c1,c2) Same analysis at the end of linac 1 (i.e., before BC2). 
(d1,d2) Same analysis at the end of BC2. 
(e1,e2) Same analysis at the end of linac 2 (i.e., before switchyard). 
The white translucent area in each panel shows the location of the nonlinear energy chip in the bunch core, which enhances the local bunch compression. 
The bunch head is at the right side in all panels.
}
\label{fig_s:Nonlinear_Chirp}
\end{figure*}

Figure~\ref{fig_s:Nonlinear_Chirp} illustrates how the nonlinear energy chirp in the bunch core developed by the double-Gaussian density modulation in the injector system is preserved throughout the accelerator up to the end of linac 2. 
Panels (a1--e1) show the simulated LPS profiles of the shaped electron beam before BC1, after BC1, before BC2, after BC2, and before the switchyard, respectively. 
To isolate the nonlinear component, panels (a2--e2) display the corresponding LPS distributions after subtracting the best-fit linear energy chirp at each location.
Once the linear term is removed, the residual nonlinear curvature in the bunch core (highlighted by the translucent white region) becomes clearly visible.

In a compression stage operated in the under-compression regime, a residual nonlinear chirp, in which particles toward the head have lower energy and those toward the tail have higher energy, produces locally enhanced bunch compression. 
This condition is consistently satisfied at all five locations as shown in simulations (Fig.~\ref{fig_s:Nonlinear_Chirp}), which demonstrates that the preserved nonlinear chirp repeatedly enhances the compression of the bunch core in BC1, BC2, and the switchyard. 
These combined effects eventually produce the few-femtosecond, 10-kA current spike in the bunch core that drives attosecond soft X-ray pulse generation in the Athos soft X-ray undulator beamline.

\subsubsection{Transverse phase space}\label{sec_SI:transverse_phase_space}

\begin{figure*}
\centering
\includegraphics[width=1\linewidth]{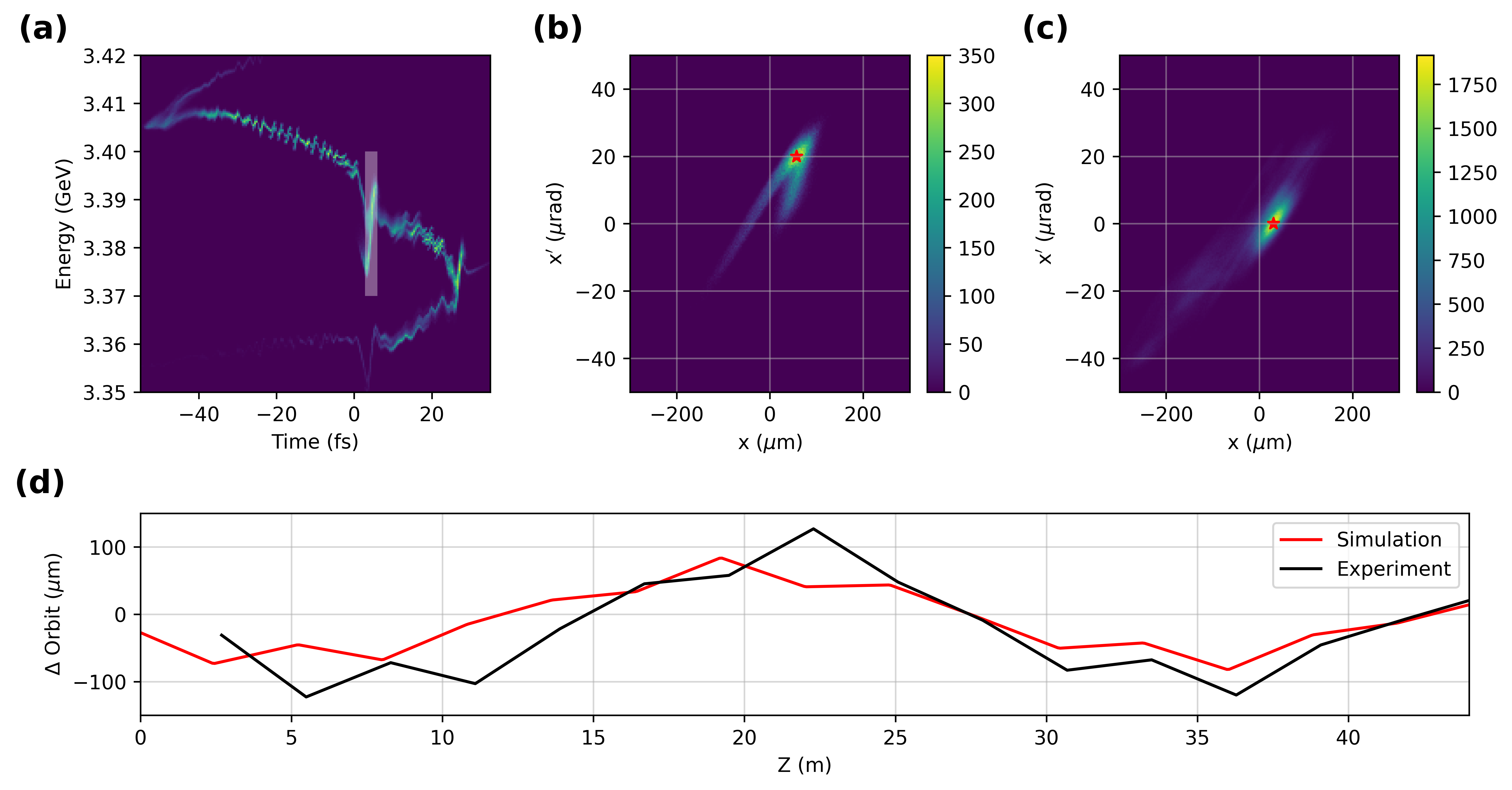}
\caption{\textbf{Time-dependent transverse phase-space.} 
Panel (a) shows the simulated electron beam LPS at the entrance of the Athos undulator beamline. 
The white translucent rectangle depicts the current spike in the bunch core. 
Bunch head is on the right side. 
Panels (b) and (c) show the simulated transverse phase-space $(x,x^{\prime})$ of electrons inside and outside the current spike, respectively. 
Red stars indicate the locations of maximum transverse phase-space density.
Panel (d) compares the simulated (red) and measured (black) differences in the horizontal central orbit along the Athos beamline when either the current spike or the pedestal current is aligned to the undulator axis. 
The measured orbit difference (black) in panel (d) corresponds to the blue curve minus the green curve in Fig.~\ref{fig:Pulse_Length_Control}g of the main text, representing the orbit difference between lasing from the current spike and lasing from the pedestal current immediately preceding the spike.
}
\label{fig_s:Transverse_Phase_Space}
\end{figure*}

Figure~\ref{fig_s:Transverse_Phase_Space} illustrates the distinct transverse dynamics of the current spike and the surrounding pedestal in the electron beam at the entrance of the Athos soft X-ray undulator beamline. 
Panel (a) shows the simulated electron beam LPS, with the shaded region marking the current spike at the bunch core.
Panels (b) and (c) show the simulated transverse phase-space $(x,x')$ of electrons inside and outside the current spike, respectively. 
The clear separation between the two distributions demonstrates that the current spike and pedestal follow distinct transverse trajectories through the undulator. 
Figure~\ref{fig_s:Transverse_Phase_Space} panel (d) further shows the difference in horizontal centroid orbits when the current spike or the pedestal current is aligned to the Athos undulator line. 
We note that the transverse lattice dispersion is closed in our simulation, so there is no significant time-dependent orbit variation along the pedestal current. 
In experiment, however, residual transverse dispersion can be nonzero. 
To highlight the CSR-induced separation of the transverse phase-space, we plot the measured orbit difference (black) obtained between the lasing configuration in which the current spike is aligned to the undulator axis and the configuration in which the longitudinal slice immediately preceding the spike is aligned instead. 
This measured orbit difference corresponds to the differential curve that is the blue curve subtracted by the green curve in Fig.~\ref{fig:Pulse_Length_Control}g of the main text. 
The good agreement between simulation (red) and measurement (black) confirms that the current spike follows a distinct transverse trajectory from the surrounding pedestal current, consistent with the time-dependent transverse phase-space separation discussed in the main text.

\subsection{Consecutive single-shot X-ray spectra}\label{sec_SI:consecutive_spectra}

\begin{figure*}
\centering
\includegraphics[width=0.8\linewidth]{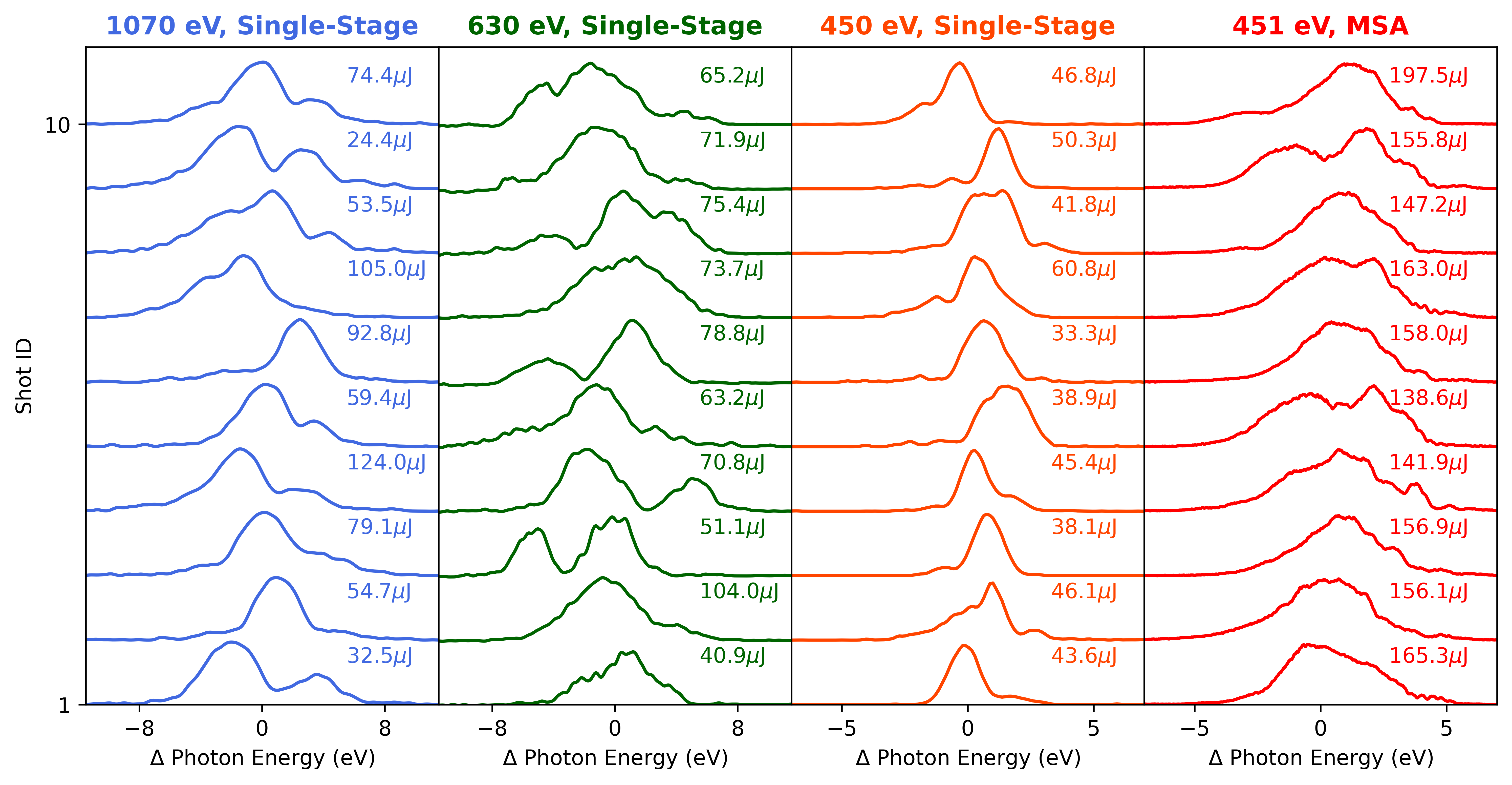}
\caption{\textbf{Consecutive single-shot spectra.}
Ten consecutive soft X-ray spectra recorded under XFEL settings optimized for the single-spike regime. 
The first three columns show X-ray pulses only generated from the current spike at 1070~eV with circular polarization (blue), 630~eV with circular polarization (green), and 450~eV with linear polarization (orange). 
The fourth column shows circularly polarized 450~eV attosecond X-ray pulses generated using the multistage amplification scheme (red). 
The corresponding pulse energy is indicated for each shot.
}
\label{fig_s:Consecutive_Shots}
\end{figure*}

Figure~\ref{fig_s:Consecutive_Shots} presents ten consecutive X-ray spectra from the same datasets, for which selected shots are presented in Figs.~\ref{fig:Exp_Data_on_XFEL_Pulses} and \ref{fig:Attosecond_MSA} in the main text. 

\subsection{Two-colour average spectrum}\label{sec_SI:two_color_average}

\begin{figure*}
\centering
\includegraphics[width=0.8\linewidth]{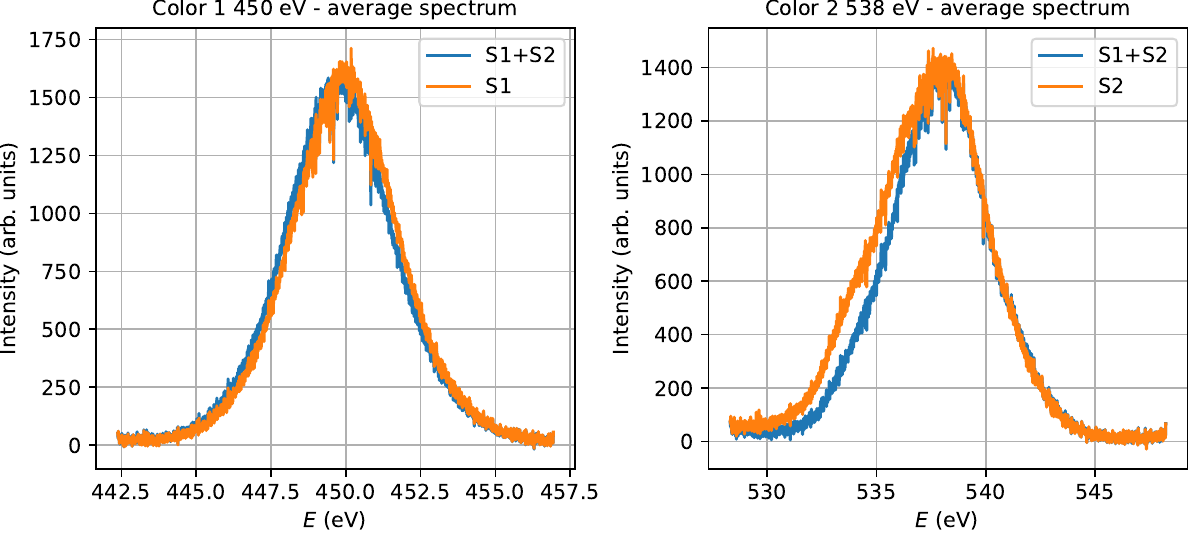}
\caption{\textbf{Independence of the two-colour lasing process.} 
Left panel: Averaged X-ray spectra at 450~eV measured when only the current spike lases (orange) and when both the current spike and pedestal current lase (blue). 
Right panel: Averaged X-ray spectra at 540~eV measured when only the pedestal current lases (orange) and when both the current spike and pedestal current lase (blue). 
The measured averaged spectra are nearly identical in both cases, demonstrating that lasing in one undulator section has negligible impact on the X-ray properties generated in the other section.
}
\label{fig_s:two_color_average}
\end{figure*}

Only one spectrometer was available during the machine study shift where the two-colour FEL pulses were generated, precluding the simultaneous acquisition of the entire FEL spectrum.
Figure~\ref{fig_s:two_color_average} shows the average spectrum of 1000~shots for the two-colour FEL setup, under the condition where with either one or both undulator stages active.  
The average spectrum centered at 450~eV is generated in the first part of the undulator beamline, and as expected, is virtually unchanged by the presence of the second pulse.
The average spectrum centered at 540~eV exhibits a bandwidth increase by about 15\% when the first undulator stage is active, which could be caused by slightly different orbit and matching conditions when the $K$ values of the initial undulator modules are detuned.
These measurements indicate independent generation of the two-colour pulses.

\subsection{Extended data on multistage amplification of attosecond X-ray pulses}\label{sec_SI:More_on_MSA}

\begin{figure*}
\centering
\includegraphics[width=1\linewidth]{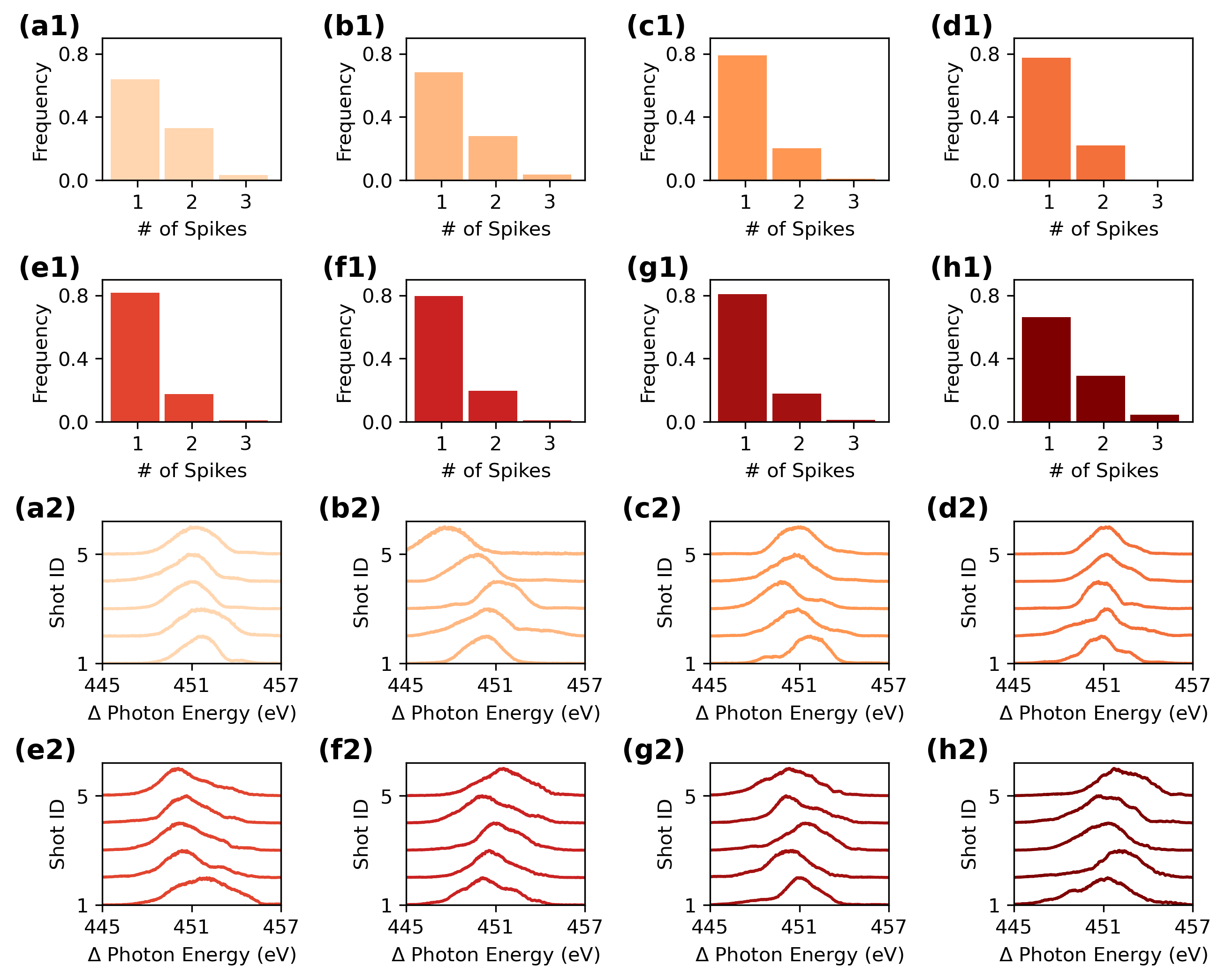}
\caption{\textbf{Extended data on multistage amplification of attosecond X-ray pulses. } 
(a1-h1) Histograms of spectral spikes counts measured with (a1) 8+0+0, (b1) 8+1+0, (c1) 8+2+0, (d1) 8+3+0, (e1) 8+4+0, (f1) 8+4+1, (g1) 8+4+2, (h1) 8+4+3 undulators enabled for X-ray lasing in the three amplification stages.
(a2-h2) Representative single-shot spectra containing a single spectral spike for the corresponding undulator configurations.
}
\label{fig_s:More_on_Attosecond_MSA}
\end{figure*}

Figure~\ref{fig_s:More_on_Attosecond_MSA} summarizes the spectral statistics of the multistage amplification (MSA) experiment for circularly polarized 450~eV attosecond X-ray pulses. 
Based on the same dataset as Fig.~\ref{fig:Attosecond_MSA} in the main text, Fig.~\ref{fig_s:More_on_Attosecond_MSA} panels (a1--h1) show the distributions of spectral-spike counts, while panels (a2-h2) present representative single-spike spectra at the end of each undulator module following the first stage. 
More than 60\% of the shots retain a single dominant spectral spike throughout the amplification process, demonstrating that the isolated-attosecond-pulse character is largely preserved during MSA. 
At the same time, the single-spike spectra progressively broaden as additional undulator modules are added to the MSA configuration, consistent with the increase in average spectral bandwidth reported in Fig.~\ref{fig:Attosecond_MSA}e.

\clearpage
\bibliography{./references.bib}
\end{document}